\documentclass[twocolumn,showpacs,prb,superscriptaddress]{revtex4}

\usepackage{amsmath}
\usepackage{tabularx}
\usepackage{graphicx}
\usepackage{color}

\begin{document}

\title{Ultrametric probe of the spin-glass state in a field}

\author{Helmut G.~Katzgraber}
\affiliation{Department of Physics and Astronomy, Texas A\&M University,
             College Station, Texas 77843-4242, USA}
\affiliation{Theoretische Physik, ETH Zurich, CH-8093 Zurich,
             Switzerland}

\author{Thomas J\"org} 
\author{Florent Krz\c{a}ka{\l}a} 
\affiliation{Laboratoire de Physico-Chimie Theorique,, UMR 
		Gulliver CNRS-ESPCI 7083, 10 rue Vauquelin, 
		75231 Paris, France}

\author{Alexander K. Hartmann}
\affiliation{Institut f\"ur Physik, Universit\"at Oldenburg, D-26111
	     Oldenburg, Germany}

\begin{abstract}

We study the ultrametric structure of phase space of one-dimensional
Ising spin glasses with random power-law interaction in an external
random field.  Although in zero field the model in both the mean-field
and non-mean-field universality classes shows an ultrametric
signature [Phys.~Rev.~Lett.~{\bf 102}, 037207 (2009)], when a field
is applied ultrametricity seems only present in the mean-field regime.
The results for the non-mean field case in an external field agree with
data for spin glasses studied within the Migdal-Kadanoff approximation.
Our results therefore suggest that the spin-glass state might be
fragile to external fields below the upper critical dimension.

\end{abstract}


\maketitle

\section{Introduction}

Spin glasses\cite{binder:86,mezard:87} are paradigmatic model
systems that find wide applicability across disciplines. Although
studied intensely, our understanding
of some of their fundamental aspects is still in its infancy.
In particular, the understanding of the nature of the spin-glass
state remains controversial and active discussion has emerged
recently.\cite{bhatt:85,ciria:93b,kawashima:96,marinari:98d,houdayer:99,krzakala:01,billoire:03b,young:04,katzgraber:05c,katzgraber:09,leuzzi:09}
It is unclear if the mean-field replica symmetry breaking (RSB)
picture\cite{parisi:79} of Parisi describes the non-mean-field behavior
of spin-glasses in an externally-applied field best.  While the
droplet theory\cite{fisher:86,fisher:87,fisher:88,bray:86}
states that there is no spin-glass state in
a field for short-range systems, the mean-field RSB
picture\cite{parisi:79,parisi:80,parisi:83,mezard:87}
states that for low enough temperatures $T$ and fields $H$ (i.e.,
below the de Almeida-Thouless line)\cite{almeida:78} a stable
spin-glass state emerges. The question lies at the core of theoretical
descriptions and is of immediate importance to applications in research
fields ranging from, for example, sociology to economics where terms linear
in the spin variable can emerge.

One way to settle the applicability of the RSB picture to short-range
spin glasses in a field while avoiding technical difficulties when
measuring observables in a field,\cite{leuzzi:09} is by
testing\cite{katzgraber:09} if the phase space is ultrametric (UM).
Unfortunately, the existence of an UM phase structure for short-range
spin glasses on hypercubic lattices remains elusive,\cite{hed:03} mainly
because only small systems can be studied numerically.  Recent results
in zero field\cite{hed:03} suggest that short-range systems are not UM,
whereas other opinions
exist.\cite{franz:00,contucci:07,contucci:08,joerg:08c}

More recently\cite{katzgraber:09} results on one-dimensional (1D) Ising
models with power-law interactions showed that short-range spin glasses
might be UM after all. Therefore, a natural probe for the spin-glass
state in a field is to study the UM response of 1D Ising models with
power-law interactions when an external field is applied.  The model has
the advantage in that by tuning the exponent of the power law, the
universality class can be tuned between a mean-field and a
non-mean-field regime. In addition, large linear system sizes can be
simulated, which allows for a better finite-size scaling analysis than
for hypercubic lattices.\cite{hed:03}

Our results show that for this model in a field the phase space has an
UM structure in the mean-field regime. However, in the non-mean-field
regime, when an external field is applied, the UM structure
seems to be much weaker for the studied system sizes, suggesting
that the spin-glass state for short-range systems is fragile with
respect to externally-applied fields. These results are compared to
studies of spin glasses within the Migdal-Kadanoff (MK) approximation.

\begin{figure}[!tbh]

\includegraphics[angle=270,width=0.9\columnwidth]{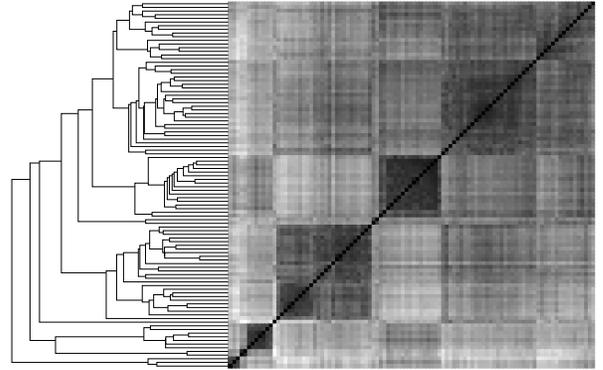}

\caption{
Dendrogram obtained by clustering 100 configurations (see text) for
a sample system with $\sigma=0.0$ (Sherrington-Kirkpatrick model)
and $L = 512$ at $T=0.36$, together with the matrix $d_{\alpha\beta}$
(grayscale, distance 0 is black). The order of the states is given by
the leaves of the dendrogram (figure rotated clockwise by $90^\circ$).
}
\label{fig:dendrogram}
\end{figure}

\section{Model}
\label{sec:model}

The 1D Ising chain with long-range power-law
interactions\cite{kotliar:83,bray:86b,fisher:88,katzgraber:03} is
described by the Hamiltonian
\begin{equation}
{\mathcal H} = - \sum_{i<j} J_{ij} S_i S_j - \sum_i h_i S_i
\; ; \;\;\;\;\;\;\;
J_{ij}= c({\sigma}) \frac{\epsilon_{ij}}{{{r_{ij}}^\sigma}} \, ,
\label{eq:model}
\end{equation}
where $S_i \in\{\pm 1\}$ are Ising spins and the sum ranges over all
spins in the system. The $L$ spins are placed on a ring to ensure
periodic boundary conditions and $r_{ij} = (L/\pi)\sin(\pi |i - j|/L)$
is the geometric distance between the spins. $\epsilon_{ij}$ are
Gaussian random couplings.  The constant $c(\sigma)$ is
chosen\cite{katzgraber:03} such that for the mean-field transition
temperature $T_c^{\rm MF}(\sigma \le 0.5, L,H=0) = 1$. In
Eq.~(\ref{eq:model}), the spins couple to site-dependent random fields
$h_i$ chosen from a Gaussian distribution with zero mean and standard
deviation $[h_i^2]_{\rm av}^{1/2} = H$.

The model has a rich phase diagram when the exponent $\sigma$ is
changed:\cite{katzgraber:03} both the universality class and the range
of the interactions can be continuously tuned.  In particular, $\sigma =
0$ gives the Sherrington-Kirkpatrick (SK) model,\cite{sherrington:75}
whose solution is the mean-field theory for spin glasses and where a
spin-glass state in a field is expected (i.e., an UM signature for low
enough $H$ and temperatures $T$).  More importantly,\cite{kotliar:83} for
$1/2 < \sigma < 2/3$ the critical behavior is mean-field-like, while for
$2/3 < \sigma \le 1$ it is non-mean-field-like.

Here we study in a field $H = 0.10$ the SK model [$\sigma = 0$] to
test our analysis protocol, as well as the 1D chain
for $\sigma = 0.60$ (also mean-field-like), as well as $\sigma
= 0.75$ ($T_c \sim 0.69$, roughly corresponding to four space
dimensions) outside the mean-field regime.  We choose two values of
$\sigma \neq 0$ to be able to discern any trends when the effective
dimensionality\cite{larson:10} is reduced. In general $d_{\rm eff} =
(2 - \eta)/(2 \sigma - 1)$, where $\eta$ is the critical exponent
for the short-range model at space dimension $d = d_{\rm eff}$.
Note that $\eta$ is zero in the mean-field regime and, for example,
$-0.275(25)$ for $d = 4$.\cite{joerg:08d}

\section{Numerical Method and Equilibration}
\label{sec:numerical}

We generate spin-glass configurations by first equilibrating the system
at low temperatures and an external random field of standard deviation
$H = 0.1$ using the parallel tempering Monte Carlo
method.\cite{geyer:91,hukushima:96} Once the system is equilibrated we
record states ensuring that these are well separated in the Markov
process and thus not correlated.  In practice, if we equilibrate the
system for $\tau_{\rm eq}$ Monte Carlo sweeps, we generate for each
disorder realization $10^3$ states separated by $\tau_{\rm eq}/10$ Monte
Carlo sweeps. We test equilibration using the method presented in
Ref.~\onlinecite{katzgraber:05c}. We consider systems sizes up to
$L=512$, which is the same maximum size as in the zero-field case
studied previously,\cite{katzgraber:09} but numerically much harder than
in the zero-field case because Monte Carlo methods equilibrate
considerably slower in a field.  For the parallel tempering simulations
$T_{\rm min} = 0.36$ and $T_{\rm max} = 1.40$ ($16$ temperatures). For
all values of $\sigma$ studied, and all system sizes $L$, we generate
$4000$ disorder realizations. For $L = 32$, the equilibration time is $2
\times 10^4$ Monte Carlo sweeps (MCS), for $64$, $1.5 \times 10^5$ MCS,
for $128$, $5 \times 10^5$, and for $256$ and $512$, $10^6$ MCS.

The presented data are for $T = 0.36$. In
Ref.~\onlinecite{katzgraber:04} we fixed $T \approx 0.4T_c$ for
all values of $\sigma$ studied to ensure that we are deep in the
spin-glass phase. However, it is unclear if one-dimensional spin
glasses with power-law interactions have a spin-glass state in a field
for $\sigma > 2/3$.\cite{katzgraber:05c,katzgraber:09b,leuzzi:09}
Using the $T_c$ estimates of Leuzzi {\em et al}.\cite{leuzzi:09}
at zero and finite field ($H = 0.1$) for the {\em diluted} version
of the model we estimate that if a spin-glass state exists for $H
= 0.1$ it should suppress the zero-field $T_c$ by approximately
20\%. For $\sigma = 0.75$ it is known that $T_c(H = 0) \approx
0.69(1)$.\cite{katzgraber:05c} Therefore $T = 0.36$ corresponds
roughly to a 40\% reduction of the critical temperature (i.e., deep
in the putative spin-glass phase).

\begin{figure}[!tbh]

\includegraphics[width=0.9\columnwidth]{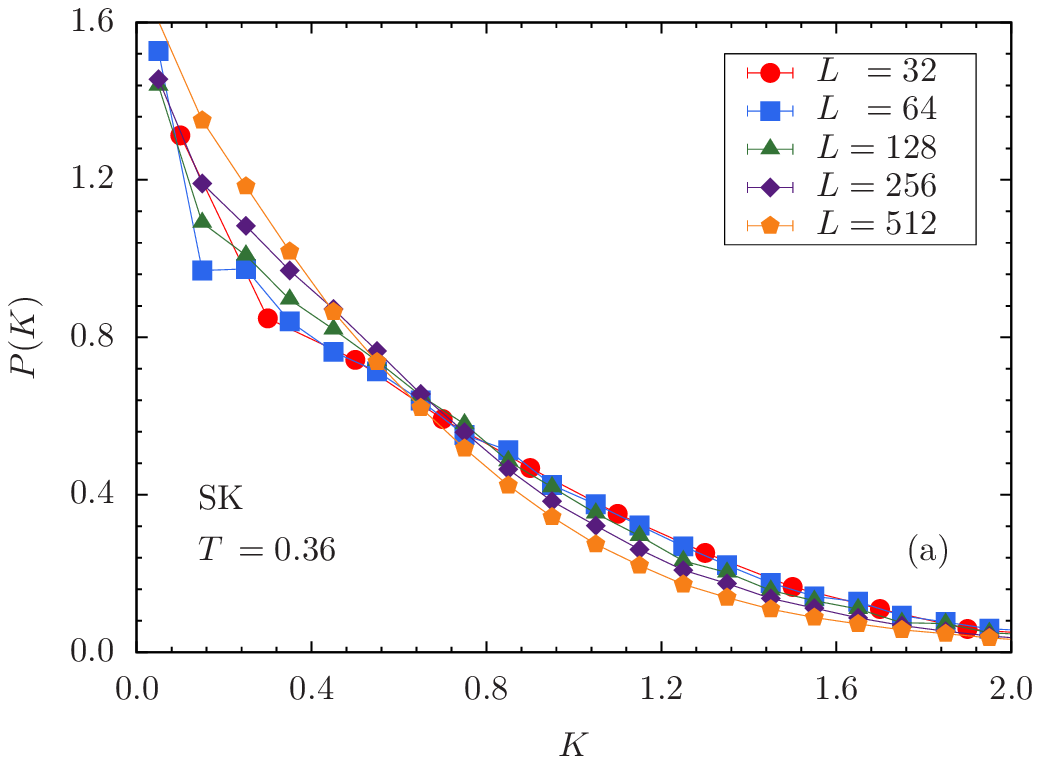}

\includegraphics[width=0.9\columnwidth]{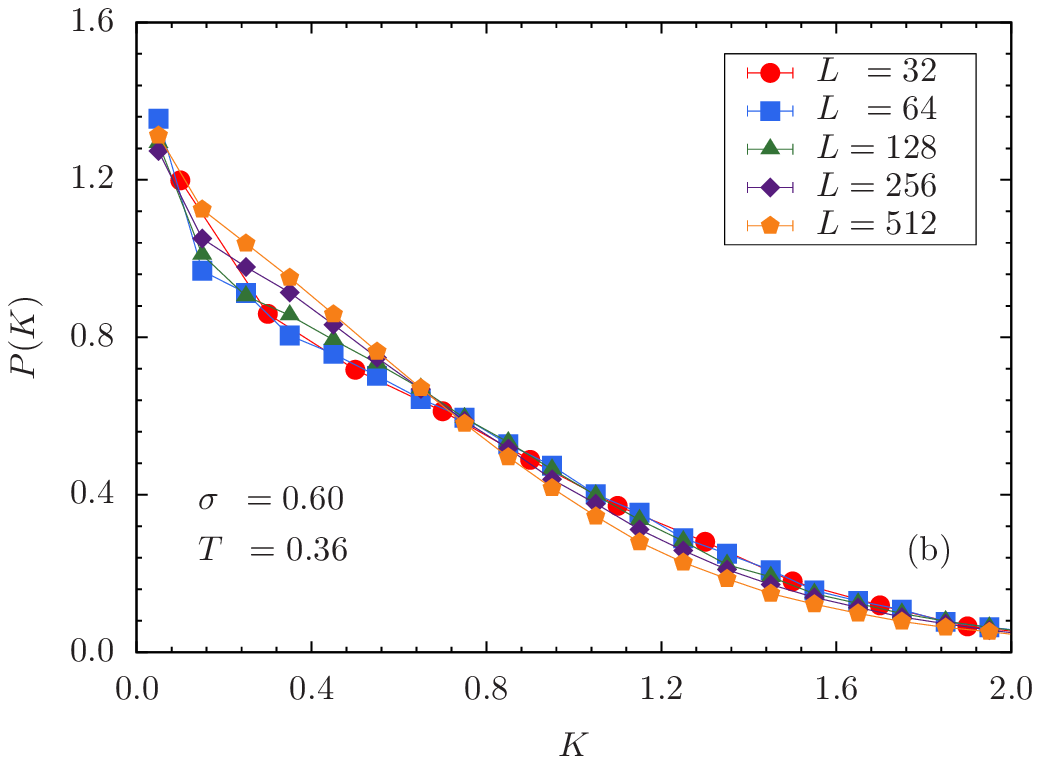}

\includegraphics[width=0.9\columnwidth]{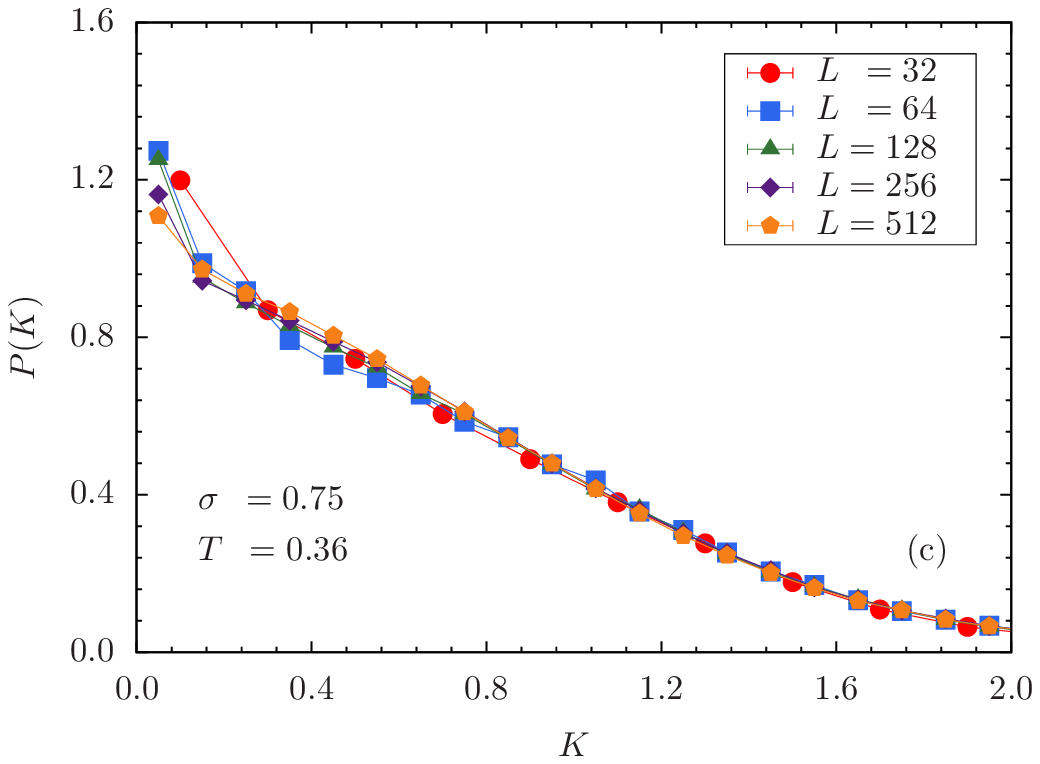}

\vspace*{-0.3cm}

\caption{(Color online)
Distribution $P(K)$ for different system sizes (all panels have the
same horizontal and vertical scale) and an external random field $H
= 0.1$.  (a) Data for the SK model. The distribution diverges very
slightly for $K \to 0$ and $L \to \infty$ thus signaling an UM phase
structure. (b) Data for $\sigma = 0.60$ (mean-field universality
class).  There is still a weak hint of a divergence for $K \to 0$.
(c) Data for $\sigma = 0.75$ (non-mean-field universality class). There
is no clear sign of a divergence in $P(K)$ for $K \to 0$. Note that when
$H = 0$ data for $\sigma = 0.75$ show a clear signature for UM
behavior.\cite{katzgraber:09} Error bars are smaller then the symbol
size.
}
\label{fig:pk}
\end{figure}

We also study spin glasses within the standard MK
approximation,\cite{kadanoff:76} (i.e., spin glasses on hierarchical
lattices).\cite{migliorini:98}  Due to the simple lattice structure,
the phase space is also expected to be simple. In fact, as shown
rigorously in Ref.~\onlinecite{gardner:84}, spin glasses on MK
lattices are replica symmetric. We used a variation of the standard
MK recursion where, starting from one bond, iteratively each bond
is replaced by $2^d$ bonds and $2^{d-1}$ spins ($d=3$). For details,
see, for example, Refs.~\onlinecite{southern:77} and \onlinecite{joerg:12}.

\section{Ultrametricity}
\label{sec:ultra}

Ultrametricity appears in different fields of research ranging
from linguistics to the taxonomy of animal species and is
a key component of Parisi's mean-field solution of the SK
model.\cite{parisi:79,mezard:84,binder:86} Therefore, if a
spin glass has no UM phase-space structure there is a strong
indication that Parisi's mean-field picture might not work for
this system.

\begin{figure}[!tbh]

\includegraphics[width=0.9\columnwidth]{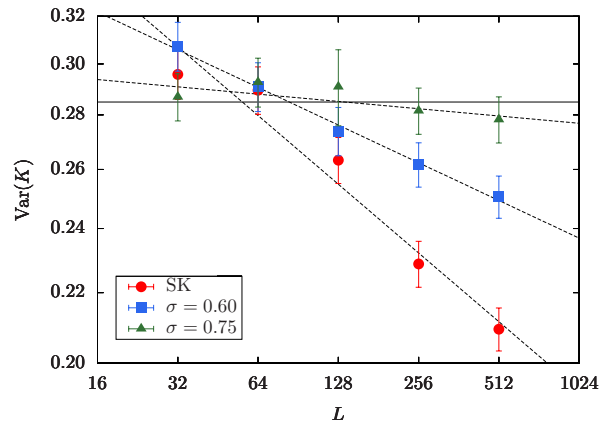}

\vspace*{-0.3cm}

\caption{(Color online)
Variance ${\rm Var}(K)$ of $P(K)$ as a function of system size $L$
for different values of $\sigma$. The data can be fit to a power law
(dashed lines).  In the mean-field regime (SK and $\sigma = 0.6$)
a fit to a constant is unlikely (see text). The power-law decay of
the variance as a function of system size suggests a divergence in
$P(K)$ for $K\to 0$.  For $\sigma = 0.75$ the data are compatible
with a constant (solid line) or a very weak power-law behavior.
}
\label{fig:variance}
\end{figure}

In an UM space\cite{rammal:86} the triangle inequality
$d_{\alpha\gamma}\le d_{\alpha\beta} + d_{\beta\gamma}$ is replaced by
a stronger condition where $d_{\alpha\gamma}\le \max\{d_{\alpha\beta},
d_{\beta\gamma}\}$ (i.e., the two longer distances must be equal and
the states lie on an isosceles triangle).  Here, $d_{\alpha\beta}$
represents the distance between two points $\alpha$ and $\beta$
in phase space.

We use the approach developed in Ref.~\onlinecite{katzgraber:09}
which is closely related to the one used by Hed {\em et al.}~in
Ref.~\onlinecite{hed:03}. For each disorder realization we produce
$M = 10^3$ equilibrium configurations.  These are sorted using the
average-linkage agglomerative clustering algorithm.\cite{jain:88}
The clustering procedure starts with $M$ clusters containing each
exactly one configuration.  Distances are measured in terms of the
Hamming distance $d_{\alpha\beta} =(1-|q_{\alpha\beta}|)$, where
$q_{\alpha\beta} = N^{-1}\sum_iS_i^{\alpha}S_i^{\beta}$ is the spin
overlap between configurations $\{S^\alpha\}$ and $\{S^\beta\}$.
Iteratively the two closest clusters $C_a$ and $C_b$ are merged
into one cluster $C_d$, reducing the number of clusters by one. The
distances of the  new cluster $C_d$ to the other remaining clusters
have to be calculated: The distance  between two clusters is the
average distance between all pairs of members of the clusters. The
iterative procedure stops when only one cluster remains, the results
are then typically structured in a tree-like structure called a
dendrogram (see Fig.\ \ref{fig:dendrogram}).  To probe for a putative
UM space structure, we randomly select three configurations from
the hierarchical cluster structure (see Ref.~\onlinecite{hed:03}),
resulting in three mutual distances. Next, we sort these Hamming
distances $d_{\rm max} \geq d_{\rm med} \geq d_{\rm min}$ and compute
$K = (d_{\rm max} - d_{\rm med})/\varrho(d)$,
where $\varrho(d)$ is the width of the distance distribution.  If the
phase space is UM, then we expect $d_{\rm max} = d_{\rm med}$ for
$L \rightarrow \infty$. Thus $P(K) \rightarrow \delta(K = 0)$ for
$L \rightarrow \infty$ and the for the variance of the distribution
${\rm Var}(K) \to 0$ for $L \rightarrow \infty$.

\begin{figure}[!tbh]

\includegraphics[width=0.9\columnwidth]{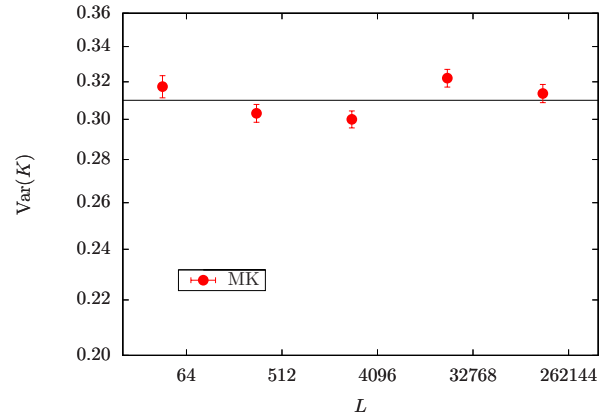}

\vspace*{-0.3cm}

\caption{(Color online)
Variance ${\rm Var}(K)$ of $P(K)$ as a function of system size $L$ for
spin glasses on MK lattices.  The data are compatible with a constant
behavior, showing that there is no UM phase-space structure for spin
glasses within the MK approximation. The solid line is a guide to the
eye.
\label{fig:MK}}
\end{figure}

\section{Results}
\label{sec:results}

Figure \ref{fig:pk}(a) shows the distribution $P(K)$ for the SK model
($\sigma = 0$), $T = 0.36$, and $H = 0.10$. There is a slight hint
for a divergence for $K\to 0$.  Similar results are found for the
mean-field regime with $\sigma = 0.60$ [Figure \ref{fig:pk}(b)].
The UM signature in a field is considerably weaker than when no
field is applied.\cite{katzgraber:09} While for the SK model there
is still a faint sign of a divergence, for larger values of $\sigma$
it is hard to see if the distributions diverge for $K \to 0$ and $L
\to \infty$.  Figure \ref{fig:pk}(c) shows data for $\sigma = 0.75$,
$T = 0.36$, and $H = 0.10$ where no clear sign of a divergence
is present, suggesting that phase space might not be UM outside the
mean-field regime.

Hence, drawing conclusions from the $P(K)$ data is not sufficient.
A better probe is given by the variance ${\rm Var}(K)$ of $P(K)$ as a
function of system size $L$ (Fig.~\ref{fig:variance}).\cite{remark:mean}
The variance of the distribution for the SK model clearly decays
with a power law ${\rm Var}(K) \sim b/L^{\gamma}$ [$b = 0.49(4)$,
$\gamma = 0.13(2)$, $Q$-factor $\sim 0.28$].\cite{press:95-ea,hartmann:09}
If we restrict the fit to $L \ge 128$ we obtain $b = 0.58(7)$ and
$\gamma = 0.16(2)$ with a $Q$-factor $\sim 0.487$. A fit to a constant
gives $Q = 0$ if the fit is performed for all data or restricted to
$L \ge 128$.  A fit to a constant+power-law behavior ${\rm Var}(K)
\sim a + b/L^{\gamma}$ gives a constant $a$ compatible with zero and a
clear power-law decay. Therefore, and as expected, the SK model shows
an ultrametric phase space structure for small externally applied
magnetic fields.

Similar results are obtained for $\sigma = 0.60$ where a fit to a power
law is very likely with $b = 0.395(6)$, $\gamma = 0.074(3)$, and $Q =
0.989$ [restricted to $L \ge 128$ we obtain $b = 0.374(1)$, $\gamma =
0.064(1)$, and $Q = 0.983$]. However, a fit to a constant gives $Q <
10^{-5}$ ($0.124$ restricted to $L \ge 128$). We also attempted a
fit to a constant+power-law behavior [i.e., ${\rm Var}(K) \sim a +
b/L^{\gamma}$]. We obtain $a = 0.18(2) > 0$ with $Q = 0.989$. This
suggests that we might be at a marginal regime (i.e., close to the
upper critical dimension).

For $\sigma = 0.75$ a fit to a very weak power law with $b = 0.30(1)$
and $\gamma = 0.014(6)$ is found with $Q = 0.897$.  Thus, the exponent
$\gamma$ is extremely small, only within about two standard deviations
from zero. Correspondingly, a fit to a constant is equally probable with
$Q = 0.811$.  Similar results are obtained for $L \ge 128$ where $b =
0.33(2)$ and $\gamma = 0.028(9)$ with $Q = 0.811$, and $Q = 0.766$ for a
fit to a constant. A fit to a constant+power-law behavior gives a
power-law exponent consistent with zero within error bars.

Summarizing, either ultrametricity in the non-mean-field regime is
completely lost in a field or greatly weakened, suggesting a marginal
signal for $\sigma = 0.60$.  Larger systems would be needed to fully
discern the behavior, however they are out of reach with current
technology. Note that for diluted systems larger system sizes are
possible, but the finite-size effects are stronger, resulting in no
overall benefit.

Within the MK approximation the distributions $P(K)$ also show no
divergence for $K \to 0$. Figure \ref{fig:MK} shows the variance of the
distributions as a function of the system size for very large lattices.
There is no discernible decrease with an increasing number of spins
(i.e., no UM structure of phase space). In fact, a fit to a power-law
behavior results in a slope compatible with zero (i.e., a constant
behavior).  This is to be expected because the model is defined on a
hierarchical lattice. However, a direct comparison to the results for
$\sigma = 0.75$ strengthens the evidence of a potential non-UM structure
for the latter case, in agreement with recent results.\cite{moore:10}

\section{Summary and Conclusion}
\label{sec:conclusions}

We have studied numerically the low-temperature configuration landscape
of long-range spin-glasses with power-law interactions.  By tuning the
exponent $\sigma$ that governs the decay of the power-law interactions
and therefore their range we can tune the system out of the mean-filed
universality class.  Using a hierarchical clustering method and
analyzing the resulting distance matrices we show that when a 
field is applied the system is only clearly UM in the mean-field regime,
unlike in the zero-field case where an UM signal was found for values of
$\sigma$ that correspond to space dimensions above and below the upper
critical dimension. Therefore, our results suggest that the spin-glass
state is fragile to an externally-applied field below the
upper critical dimension. Larger systems would be needed to determine if
the UM signature for $\sigma = 0.75$ (corresponding approximately to
four space dimensions) persists in a field or not.

\begin{acknowledgments}

H.G.K.~acknowledges support from the Swiss National Science Foundation
(Grant No.~PP002-114713) and the National Science Foundation (Grant
No.~DMR-1151387). We thank Texas A\&M University, the Texas Advanced
Computing Center (TACC) at The University of Texas at Austin, the
Centro de Supercomputaci{\'o}ny Visualizaci{\'o}n de Madrid (CeSViMa)
and ETH Zurich for HPC resources.

\end{acknowledgments}

\bibliography{refs,comments}

\begin{thebibliography}{48}
\bibitem[{rem()}]{remark:mean}
\bibinfo{note}{The mean of the distribution also decreases with system size in
  the mean-field regime. This is not the case outside the mean-field regime
  ($\sigma = 0.75$).}
\end{thebibliography}

\end{document}